\documentclass[preprint,preprintnumbers,amsmath,amssymb,nofootinbib,pra]{revtex4}
\usepackage{amsfonts}
\usepackage{amssymb}
\usepackage{latexsym}
\usepackage{graphicx}
\usepackage{longtable}
\usepackage[active]{srcltx}
\newcommand{\al}{\alpha}

\newcommand{\De}{\Delta}

\newcommand{\lrar}{\leftrightarrow}

\begin{document}


\title{A note about the ground state of the ${\rm H}_3^+$ hydrogen molecular ion}

\author{J. C. Lopez Vieyra}
\email{vieyra@nucleares.unam.mx}
\author{A.V.~Turbiner}
\email{turbiner@nucleares.unam.mx}

\author{H. Medel}
\email{medel@nucleares.unam.mx}

\affiliation{Instituto de Ciencias Nucleares, Universidad Nacional
Aut\'onoma de M\'exico, Apartado Postal 70-543, 04510 M\'exico,
D.F., Mexico}

\date{May 25, 2011}

\begin{abstract}
Three simple $7-, (7+3)-, 10-$parametric trial functions for the ${\rm H}_3^+$ molecular ion are presented. Each of them provides subsequently the most accurate approximation for the  Born-Oppenheimer ground state energy among several-parametric trial functions. These trial functions are chosen following a criterion of physical adequacy and includes the electronic correlation in the exponential form $\sim\exp{(\gamma r_{12})}$, where $\gamma$ is a variational parameter. The Born-Oppenheimer energy is found to be $E=-1.340 34, -1.340 73, -1.341 59$\,a.u., respectively, for optimal equilateral triangular configuration of protons with the equilibrium interproton distance $R=1.65$\,a.u. The variational energy agrees in three significant digits (s.d.) with most accurate results available at present as well as for major expectation values.
\end{abstract}
%

\maketitle

\section{Introduction}

The two-electron hydrogenic molecular ion ${\rm H}_3^+$ is one among the most abundant chemical compounds  in the Universe. Its existence is of fundamental importance  in  chemistry and physics, in particular, because of its stability towards decay to ${\rm H}_2 + p$, the ${\rm H}_3^+$ ion is also a major proton donor in chemical reactions in interstellar space.  The ${\rm H}_3^+$ was discovered experimentally by J.J.~Thomson in 1912~\cite{Thomson:1912}. The system was very difficult for theoretical studies. Many theoretical methods were developed to study low-lying quantum states of this system. In particular, it became clear very quickly that interelectron correlation is of great importance and it should be included to the variational trial function explicitly which assure a faster convergence. This conclusion was similar to one drawn by James and Coolidge for the ${\rm H}_2$ molecule. Usually, the interelectron correlation was written in the form $r_{12}^n$ (Hylleraas~\cite{Hylleraas:1930} - James-Coolidge~\cite{JC:1933} form) or $\exp(-\alpha
 r_{12}^2)$ (Gaussian form, see e.g. Ref.~\cite{Kutzelnigg:2006}).

Recently, Korobov~\cite{Korobov:2000} showed in explicit way that for the case of Helium atom the use of exponential form $\exp(-\gamma r_{12})$ dramatically improves convergence and leads, in fact, to the most accurate results for the ground state energy for the Helium atom at present. Later on, it was shown that the similar use of exponential correlation  $\exp(-\gamma r_{12})$ for the ${\rm H}_2$ molecule allows to construct the most accurate trial function among few-parametric trial functions~\cite{TG:2007}.
A hint why namely this $r_{12}$-dependence leads to the fast convergent results
was given in \cite{TG:2007}. In year 2006 an overwhelming discussion meeting took
place in London, UK where different properties of the  ${\rm H}_3^+$ ion and, in particular, various theoretical approaches to study the ${\rm H}_3^+$ ion were
exposed (see \cite{Oka:2006}).

The goal of this Note is to propose a simple, compact, easy-to-handle trial function depending exponentially on $r_{12}$  with few nonlinear parameters which leads to highly accurate Born-Oppenheimer ground state energy and major expectation values. We are not aware about previous studies of the ${\rm H}_3^+$ ion with trial functions involving $r_{12}$ in exponential form with a single exception \cite{2007} where the ${\rm H}_3^+$
in linear configuration was explored.

In this paper atomic units ($\hbar=e=m_e=1$) are used throughout, albeit energies are
given sometimes in Rydbergs.

\section{The ${\rm H}_3^+$ ion in the Born-Oppenheimer approximation}

\begin{figure}[tb]
\begin{center}
   \includegraphics*[width=4.in,angle=0.0]{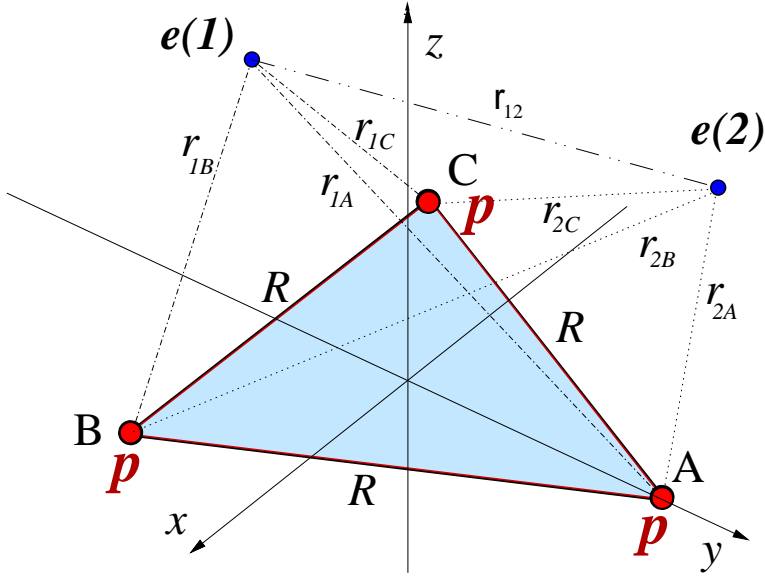}
    \caption{\label{fig1} Geometrical setting  for the hydrogen molecular ion ${\rm H}_3^+$ in equilateral triangular configuration.  The three protons are located on the $x$-$y$ plane
    forming an equilateral triangle with the origin of coordinates located at the geometrical center  (circumcenter) of the triangle.}
\end{center}
\end{figure}

The Hamiltonian which describes the  ion ${\rm H}_3^+$  under the assumption that the protons are infinitely massive (the
Born-Oppenheimer approximation of zero order) and located at the vertices of an equilateral triangle of side $R$
(see Fig.~\ref{fig1} for the geometrical setting and notations), is written as
follows:
\begin{equation}
\label{H}
  {\cal H}\ =\sum_{j=1}^2 {\hat {\mathbf p}_{j}}^2\ -
  \ \sum_{\buildrel{{j}=1,2}\over{\kappa =A,B,C}}
  \frac{2}{r_{{j},\kappa}}
  \ +\ \frac{2}{{\sf r}_{12}}\ +\ \frac{6}{R}  \ ,
\end{equation}
where ${\hat {\mathbf p}_{j}}=-i \nabla_{j}$ is the 3-vector
of the momentum of the ${j}$th electron, the index $\kappa$ runs
over protons $A$, $B$ and $C$, $r_{{j},\kappa}$ is the distance
between the ${j}$th electron and the $\kappa$th proton, ${\sf r}_{12}$ is the
interelectron distance, and  $R$ is the interproton distance.

It is a well established fact that the ground state of the ${\rm H}_3^+$ molecular ion
is $1 {}^1 A_1^\prime$, an electronic  spin-singlet state, with the three protons forming
an equilateral triangle in the totally symmetric  representation $A_1^\prime$  of a
$D_{3h}$ point symmetry \cite{Tennyson:1995}. Thus, the ground state electronic wavefunction
should be symmetric under permutations of the three indistinguishable protons. This ground state
is the major focus of the present study.


%
%

It is worth  mentioning that the best theoretical value at the moment for the Born-Oppenheimer
ground state energy is \hbox{$E=-1.343 835 625 02$\,a.u.} \cite{Pavanello:2009} obtained
with a basis of 1000 explicitly correlated spherical Gaussian functions with
shifted centers. This value surpasses the previous record $E=-1.343 835 624$\,a.u.
by Cencek {\it et al.} which was obtained by using explicitly correlated Gaussian
functions \cite{Cencek:1995}.

\section{Variational method}

The variational procedure is used as a method to explore the problem. The recipe of
choosing the trial function is based on arguments of physical relevance, e.g. the trial
function should support the symmetries of the system, has to reproduce the Coulomb
singularities and the asymptotic behavior at large distances adequately
(see, e.g. \cite{turbinervar, turbinervar1, Turbiner:2006}).
In practice, the use of such trial functions implies the convergence of a
special form of the perturbation theory where the variational energy
is the sum of the first two terms. Let us remind the essentials of
this perturbation theory (for details, see \cite{turbinervar, turbinervar1, Turbiner:2006}). Let
us assume that our original Hamiltonian has a form ${\cal H}=-\Delta
+ V$. As a first step we choose a trial function $\psi^{(trial)}$
(normalized to one) and find a potential for which such a
trial function $\psi^{(trial)}$ is an exact eigenfunction, {\it i.e.}
$V_{trial}={\De \psi^{(trial)}}/{\psi^{(trial)}}$, with
energy $E_{trial}=0$. In a pure formal way we can construct a
Hamiltonian ${\cal H}_{trial} = -\Delta + V_{trial}$ such that
${\cal H}_{trial} \psi^{(trial)}=0$. It can be easily shown that the
variational energy
\[
 E_{var}= \langle\psi^{(trial)}|{\cal H}|\psi^{(trial)}\rangle
\]
is nothing but the first two terms in the perturbation theory where
the unperturbed problem is given by ${\cal H}_{trial}$ and the
perturbation is the deviation of the original potential $V$ from the
trial potential $V_{trial}$, namely, $V_{perturbation}=V-V_{trial}$.
Eventually, we arrive at the formula
\begin{equation}
    E_{var}= E_{trial} + E_1 (V_{perturbation})\ ,
\end{equation}
here $E_1
(V_{perturbation})=\langle\psi^{(trial)}|V_{perturbation}|\psi^{(trial)}\rangle$
is the first energy correction in the perturbation theory, where
the unperturbed potential is $V_{trial}$. It is worth noting that if the
trial function is the Hartree-Fock function the resulting
perturbation theory is nothing but the Moeller-Plesset perturbation
theory (see, e.g. \cite{Levine}, Section 15.18)\footnote{It is
worth noting that the question about a convergence of the Moeller
Plesset perturbation theory is not settled yet \cite{MP}}.

One of the criteria of convergence of the perturbation theory in
$V_{perturbation}=V-V_{trial}$ is a requirement that the ratio
$|{V_{perturbation}}/{V}|$ should not grow when $r$ tends to
infinity in any direction. If this ratio is bounded by a constant it
should be less than one. In fact, it is a condition that the
perturbation potential is subordinate with respect to the
unperturbed potential. The value of this constant controls the rate of
convergence - a smaller value of this constant leads to faster
convergence \cite{turbinervar1}. Hence, the above condition gives a
importance to the large-range behavior of the trial functions. In
the physics language the above requirement means that the phenomenon
of the Dyson's instability should not occur (for a discussion see
\cite{turbinervar}) \footnote{It is worth noting that this procedure
for a selection of the trial function was applied successfully to a
study of one-two-electron molecular systems in a magnetic field leading
to the highly accurate results. Many of these results are the most
accurate at the moment (see \cite{Turbiner:2006} and \cite{Turbiner:2010}).}.

\section{Correlated trial function}

Among different forms to include explicit electronic correlation in the
trial wave function for two-electron problems we mention three major
approaches (see e.g. \cite{Klopper:2006}):
the linear in $r_{12}$, the gaussian $\exp(-\alpha r_{12}^2)$ and
exponential $\exp(\gamma r_{12})$ terms. Among them, the factor
$\exp(\gamma r_{12})$ only  fulfills the adequacy requirements for a trial
function described above. Thus, following the  guidelines of
Section III and the requirement of the convergence of the   perturbation theory,
we choose the trial function for the ground state in the following form:
\begin{equation}
\label{ansatzG}
 \psi_{0} = (1+ P_{12}) \sum_{{\rm perm}\{ A,B,C\}}
 e^{
 -\al_1 r_{1A}-\al_2 r_{1B} -\al_3 r_{1C} -\al_4 r_{2A}
 -\al_5 r_{2B} -\al_6 r_{2C}
 + \gamma r_{12} }\, ,
 \end{equation}
where the sum runs over the permutations of the identical protons $A,B,C$ ($S_3$ symmetry), and $P_{12}$ is the operator which interchanges electrons $(1 \lrar 2)$. The variational parameters consist of non-linear parameters $\al_{1-6}$ and $\gamma$ which characterize the (anti)screening of the Coulomb charges. The interproton distance $R$, see Fig.1 is kept fixed. It is chosen to be equal $R=R_{eq}=1.65$a.u [10-11]. The function (\ref{ansatzG}) is a symmetrized product of $1s$ Slater type orbitals multiplied by the exponential correlation factor $e^{\gamma r_{12}}$.

Calculations of the variational energy were performed using the minimization package MINUIT from CERN-LIB. Six-dimensional integrals
which appear in the functional of energy were calculated numerically using a "state-of-the-art" dynamical partitioning procedure:
the domain of integration was divided into 972 subdomains following the profile of the integrand, in particular, separating out
the domains with sharp changes of the integrand. Then each subdomain was integrated separately in parallel manner with controlled
accuracy (for details, see e.g. \cite{Turbiner:2006}). A realization of the routine required a lot of attention and care. During
the minimization process the partitioning was permanently controlled and adjusted. Numerical integration of every subdomain was
done with a relative accuracy \hbox{of $\sim 10^{-3} - 10^{-7}$} depending on its complexity and relative contribution using an
adaptive routine based on an algorithm by Genz and Malik~\cite{GenzMalik} from R. Sch\"urer's HIntLib C++
multidimensional integration library (http://mint.sbg.ac.at/rudi/). Parallelization was reached using the MPI standard
library MPICH. Computations were performed on a Linux cluster with 48 Xeon processors of 2.67\,GHz each, and 12Gb total RAM
plus extra processor serving as the master node. Total minimization process took about 1000 hours of wall clock time when a single
call took about three minutes. For optimal values of parameters it took about 20 minutes  (wall clock time) to compute
a variational energy with relative accuracy $10^{-7}$.

\section{Results}

In Table~\ref{table1} we present the results for the ground state energy at interproton
equilibrium distance of the ${\rm H}_3^+$ molecular ion obtained by different researchers using different methods.
In a clear way it is seen that the Born-Oppenheimer ground state energy obtained using the trial function (\ref{ansatzG})
is the most accurate (the lowest) energy obtained with a few parametric functions. In particular, the trial function (\ref{ansatzG})
gives a lower energy than the energies obtained with the explicitly correlated functions based on both Gaussians in
$r_{12}$ \cite{Salmon:1973} and linear in $r_{12}$ \cite{Klopper:1993}, when a relatively small number of terms with
non-linear parameters is involved. The trial function (\ref{ansatzG}) is more accurate than almost all(!) traditional
CI calculations which were performed before 1971 (see \cite{Anderson:1992}) even including one of the largest set of 100
configurations \cite{Kawaoka:1971}. In those CI calculations no explicit correlation was included. The variational energy
obtained with (\ref{ansatzG}) is even of comparable accuracy to the large CI calculations \cite{DubenII:1971, Kawaoka:1971}
\footnote{for a list of 42 calculations of the ground state energy of ${\rm H}_3^+$ in the period 1938-1992 see
Ref.\cite{Anderson:1992}, for a list of selected {\it ab-initio} calculations till 1995, see \cite{Kutzelnigg:2006}}.
Table~\ref{table1b} shows the optimal values of the variational parameters in (\ref{ansatzG}).

The list of major expectation values obtained with the trial function (\ref{ansatzG})
and its comparison with results of other calculations is given in Table~\ref{table2}.
A reasonable agreement for expectation values is observed. In particular, for the expectation values $\langle 1/r_{1 A}\rangle$, $\langle x^2\rangle$,
$\langle z^2\rangle$ and $\langle r^2\rangle$  an agreement within
$\sim$ 1\% with ours and all other calculations is observed, including ones obtained in the large CISD-R12 calculations \cite{Klopper:1993}.
Also, for the expectation value of $\langle 1/r_{12}\rangle$ we have an agreement in the first significant digit with other calculations being in closer agreement to the value obtained with the correlated Gaussian (unrestricted) wavefunction with 15 terms \cite{Salmon:1973}, while for the expectation values $\langle 1/r_{1 A}\rangle$ and $\langle z^2 \rangle$ we observe an agreement with other calculations in 3 and 2 significant digits, respectively. These facts seem to indicate that the presented expectation values are very accurate, corroborating the quality of the trial function (\ref{ansatzG}) giving 2-3 s.d. correctly. Perhaps, it is worth noting that in absence of any criteria about accuracy of the obtained expectation values we can only note about agreement of them obtained in different approaches.

%
%

\begin{table}
\begin{tabular}{|l|l|l|l|}
\hline
\hspace{10pt} $E$ (a.u.) & $R$ (a.u.) &  \hspace{40pt} method & reference \\ \hline \hline
 -1.339 7  & 1.66  &  CI-GTO,  $>$ 120 configs  & \cite{Ciszmadia:1970}   (1970)\\ \hline
 -1.306 29  & 1.65     &   GG, 3 terms, 5 non-linear params   &
   \cite{Salmon:1973} (1973) \\
 -1.327 25  & 1.65     &   GG, 6 terms, 7 non-linear params   &
     \\
 -1.331 47  & 1.65     &   GG, 10 terms, 9 non-linear params   &
      \\
 -1.332 29  & 1.65     &   GG, 15 terms, 11 non-linear params   &
      \\  \hline
 -1.334 382 &  1.65  &  R12, $10s$ basis set  &  \cite{Klopper:1993} (1993)  \\
 -1.334 632 &  1.65  &  R12, $30s$ basis set  &     \\ \hline
 -1.340 34  &  1.65  &  7-Parametric Trial Function (\ref{ansatzG}) & present \\
 \hline
 -1.340 5   & 1.6405 &   CI -GTO, 48 configs   &   \cite{DubenII:1971}  (1971)\\ \hline
 -1.340 5   & 1.65   &   CI -STO, 100 configs  &   \cite{Kawaoka:1971} (1971)\\ \hline
 -1.340 73  & 1.65   & (7+3)-Parametric Trial Function (\ref{ansatzGHL1}) ${}^{(i)}$   & present \\[2pt]
 -1.341 59  & 1.65   & 10-Parametric Trial Function    (\ref{ansatzGHL1}) ${}^{(ii)}$  & present \\ \hline
-1.342 72   & 1.65041 & CI-GTO, 108 terms & \cite{Burton:1985} (1985)\\ 
%
 -1.342 784 & 1.6504 & CI-GTO,  $8s3p1d/[6s3p1d]$ basis set& \cite{Dykstra:1978} (1978)\\
-1.343 40   & 1.6504  & CI-GTO,  $10s4p2d$ basis set & \cite{Meyer:1986} (1990)\\
-1.343 822  & 1.65   & CI-GTO, 700 terms& \cite{Alexander:1990} (1990)\\
\hline
-1.342 03 & 1.6504  & CI with r12, 36 configs & \cite{Preiskorn:1982} (1982)\\
-1.343 422 & 1.6504  & CI with r12, 192 configs & \cite{Preiskorn:1984} (1984)\\
-1.343 500 &  1.6504  & CI with r12, $13s3p/[10s2p]$ basis set & \cite{Urdaneta:1988} (1988)\\ 
-1.343 828  & 1.65  & CI with r12,  $13s5p3d$ basis set. & \cite{Frye:1990} (1990)\\ \hline
 -1.343 835  & 1.65   &  R12, $30s20p12d9f$ basis set  &  \cite{Klopper:1993} (1993)  \\ \hline
 -1.343 35  & 1.65    &   GG, 15 terms, 135 non-linear params   &
   \cite{Salmon:1973} (1973) \\ 
 -1.343 835 624  &  1.65  &  GG,  600 terms  &  \cite{Cencek:1995} (1995)  \\ \hline
 -1.343 835 625 02 & 1.65 & ECSG, 1000 terms & \cite{Pavanello:2009} (2009)\\
\hline
\end{tabular}
\caption{\label{table1}
 A selection of the calculations for the Born-Oppenheimer ground state energy at equilibrium distance of ${\rm H}_3^+$. Record calculations of Ref.~\cite{Pavanello:2009} (2009) and Ref.\cite{Cencek:1995} (1995). CI denotes Configuration Interaction, STO - Slater Type Orbitals, GTO - Gaussian Type Orbitals, GG - correlated Gaussians (Gaussian Geminals), R12 - the CI calculation augmented by terms linear in $r_{12}$, ECSG - Explicitly Correlated Spherical Gaussian functions.
 ${}^{(i)}$ Trial Function (\ref{ansatzGHL1}) with the parameters of $\psi_{0}$ kept fixed and equal to ones found for (\ref{ansatzG}), ${}^{(ii)}$ Trial Function (\ref{ansatzGHL1}) with all 10 parameters optimized.}
\end{table}


\begin{table}
\begin{tabular}{|c|ccccccc|ccc|}
\hline
$E$ (Ry)   & $\al_1$  & $\al_2$ & $\al_3$ & $\al_4$ & $\al_5$ & $\al_6$ & $\gamma$ & $A$ & $\tilde\al $ & $\tilde\gamma$  \\ \hline \hline
-2.680 7   & -0.00353 & 0.18548 & 1.4245  & 1.0471  & 0.15082 & 0.58912 & 0.21632  &   --  &      --        &      --           \\
\hline
-2.681 4   &    \multicolumn{7}{|c|}{${}^{\prime\prime}\qquad\quad$ }              & -0.03000    &  0.47517 &  0.76398     \\
\hline\hline
-2.683 2  & -0.00294 & 0.21022 & 1.3849 & 1.0199 & 0.17103 & 0.59084  & 0.26044 &   -0.51154   &  0.59589  &  0.86229     \\
\hline\hline
\end{tabular}
\caption{\label{table1b}
  The ground state energy of ${\rm H}_3^+$ at \hbox{$R_{eq}=1.65$\,a.u.} and the non-linear variational parameters in $[a.u.]^{-1}$ corresponding to the trial function (\ref{ansatzG}), to the trial function (\ref{ansatzGHL1}) with parameters corresponding to $\psi_0$ fixed and to (\ref{ansatzGHL1}) with 10 optimized parameters.}
\end{table}

\begin{table}[htdp]
\begin{center}
\begin{tabular}{|l|c|c|}
\hline
Expectation Value & Trial Function
\raise 2pt \hbox{(\ref{ansatzG})} / \hbox{(\ref{ansatzGHL1})${}^{(i)}$}  / \lower 2pt \hbox{(\ref{ansatzGHL1})${}^{(ii)}$}  &  Others\\
\hline
$\langle r_{12}\rangle$                  & 2.0032      &    \\[-5pt]
                                         & 2.0013      &    \\
                                         & 1.9931      &    \\
$\langle 1/r_{12}\rangle$                & 0.6315      & 0.59415${}^a$ \\[-5pt]
                                         & 0.6304      & 0.62636${}^c$ \\
                                         & 0.6302      &    \\
$\langle 1/r_{1 A}\rangle$               & 0.8548      & 0.85519${}^c$ \\[-5pt]
                                         & 0.8548      & 0.8553${}^e$ \\
                                         & 0.8549      &    \\
$\langle x^2\rangle =\langle y^2\rangle$
                                         & 0.7711 & 0.75818${}^a$ \\[-5pt]
                                         & 0.7703 & 0.7595${}^b$ \\
                                         & 0.7666 & 0.75913${}^c$ \\
                                         &        & 0.75968${}^d$ \\
                                         &        & 0.7605${}^e$   \\
$\langle z^2\rangle$                     & 0.5399 & 0.54802${}^a$  \\[-5pt]
                                         & 0.5367 & 0.5451${}^b$  \\
                                         & 0.5337 & 0.54085${}^c$ \\
                                         &        & 0.54179${}^d$ \\
                                         &        & 0.5396${}^e$   \\
$\langle r^2\rangle$                     & 2.0822 & 2.06442${}^a$\\[-5pt]
                                         & 2.0773 & 2.0640${}^b$\\
                                         & 2.0669 & 2.05911${}^c$ \\
                                         &        & 2.06114${}^d$ \\
\hline
\end{tabular}
\end{center}
\caption{\label{table2}
   Expectation values (in a.u.) for the ${\rm H}_3^+$ ion in its ground state obtained
   with the trial functions (\ref{ansatzG}) and (\ref{ansatzGHL1})${}^{(i,ii)}$. Corresponding results obtained with other methods are displayed for comparison. Coordinates $x,y,z$  and $r$ are measured from the center of the equilateral triangle formed by protons.
   ${}^a$ Ref.\cite{DubenII:1971} CI-43;
   ${}^b$ CI wavefuncion (I) in Ref.\cite{Kawaoka:1971};
   ${}^c$ Correlated Gaussian (unrestricted) wavefunction with 15 terms  in Ref.\cite{Salmon:1973};
   ${}^d$ CI wavefunction in Ref.\cite{Carney:1974};
   ${}^e$ CISD-R12 wavefunction with the $10s8p6d4f$  basis set in Ref.\cite{Klopper:1993}.
}
\end{table}%

\section{Conclusion}

We presented a simple and compact 7-parametric variational trial function together with its possible natural generalization by addition of the Heitler-London (HL) type function. This function already provides surprisingly accurate Born-Oppenheimer energy for the ground state of such a complicated molecular system ${\rm H}_3^+$. It is chosen following a criterion of physical adequacy which suggests to take the electronic correlation in the form $\sim\exp{(\gamma r_{12})}$ where $\gamma$ is a variational parameter. The minimum energy is found to be $E=-1.340 34$\,a.u. at an equilibrium interproton distance $R=1.65$\,a.u.  This result for the energy is the most accurate among the values obtained with several parametric trial functions. In particular, it is more accurate than the energies obtained with the explicitly correlated approaches of Ref.\cite{Klopper:1993} (linear in $r_{12}$) and that of Ref.\cite{Salmon:1973} (Gaussian in $r_{12}$), when a relatively small number of terms
and non-linear parameters are involved.

In a spirit of the approach presented in \cite{TG:2007} (see also \cite{Korobov:2000}) the trial function (\ref{ansatzG}) can be modified by adding similar function, in particular, of the Heitler-London type:
\begin{equation}
\label{HL}
 \psi_{HL}\ =\
 e^{-\tilde\al ( r_{1A}-  r_{1B} -   r_{1C} -  r_{2A} -  r_{2B} -  r_{2C})
 + \tilde\gamma r_{12} }\, ,
\end{equation}
where $\tilde \al, \tilde \gamma$ are parameters. The function (\ref{HL}) alone gives a dominant contribution to small interproton distances. Taking a linear superposition with (\ref{ansatzG})
\begin{equation}
\label{ansatzGHL1}
 \Psi = \psi_{0} +  A\, \psi_{HL} \,,
\end{equation}
and making minimization with respect to parameters $A, \tilde\alpha$ and $\tilde\gamma$ only (see Table \ref{table1b}) give an essential improvement in the energy (see Table \ref{table1}). In particular, this function, which contains $(7 + 3)$ variational parameters, allows us to get more accurate result for energy than one obtained in \cite{Kawaoka:1971} within CI-STO with 100 configurations.

Releasing all 10 parameters in (\ref{ansatzGHL1}) (see Table \ref{table1b}) we obtain further improved result (see Table \ref{table1}), although we are still unable to reproduce the fourth significant figure in the energy. However, the obtained energy is among the thirteen the most accurate variational results ever calculated so far. It is slightly 
worse then one obtained in \cite{Preiskorn:1982} based on CI with r12 method with 36 configurations. The expectation values in Table \ref{table2} gradually change with move from one Ansatz to another seemingly demonstrating a convergence. It seems evident that taking a linear superposition of two (or more) functions (\ref{ansatzG})  instead of (\ref{ansatzGHL1}) will improve essentially the variational energies. It will be done elsewhere. Undoubtfully, trial functions (\ref{ansatzG}), (\ref{ansatzGHL1}) can be used to study potential energy surface.
It is worth emphasizing that the main attraction of functions (\ref{ansatzG}), (\ref{ansatzGHL1}) is their compactness.

The function (\ref{ansatzG}) can be easily modified for a study of spin-triplet states
and as well as the low-lying states with non-vanishing magnetic quantum number. A generalization to more-than-two electron molecular systems seems also straightforward.

\section*{Acknowledgement}

This work was supported in part by PAPIIT grant {\bf IN115709}
(Mexico). The authors are deeply thankful to D. Turbiner for a help
with computer code design, for a creation of the optimal configuration
for 48-processor cluster used for calculation. The authors are also obliged to E. Palacios
for his technical support with the cluster {\it Karen}.
The second author is grateful to the University Program
FENOMEC (UNAM, Mexico) for a support.

\end{document}